\documentclass[pdflatex,sn-mathphys-num]{sn-jnl}

\usepackage{graphicx}%
\usepackage{multirow}%
\usepackage{amsmath,amssymb,amsfonts}%
\usepackage{amsthm}%
\usepackage{mathrsfs}%
\usepackage[title]{appendix}%
\usepackage{xcolor}%
\usepackage{textcomp}%
\usepackage{manyfoot}%
\usepackage{booktabs}%
\usepackage{algorithm}%
\usepackage{algorithmicx}%
\usepackage{algpseudocode}%
\usepackage{listings}%
\usepackage{setspace}%
\usepackage{geometry}
\usepackage[switch,running]{lineno}  

\begin{document}

\title[Kerr-induced In-line Interferometry]{Kerr-induced In-line Interferometry for Ultra-sensitive Phase-Contrast Imaging}

\author[1]{\fnm{Glitta R.} \sur{Cheeran}}\email{glittarosalia.cheeran@leibniz-ipht.de}
\author[1]{\fnm{Mehmet} \sur{M\"uft\"uoglu}}
\author[1]{\fnm{Sobhi} \sur{Saeed}}
\author[1]{\fnm{Bennet} \sur{Fischer}}
\author*[1,2]{\fnm{Mario} \sur{Chemnitz}}\email{chemnitzmario@ipht-jena.de}

\affil[1]{\orgname{Leibniz Institute of Photonic Technology}, \orgaddress{\street{Albert-Einstein-Str.9}, \city{Jena}, \postcode{07745},  \country{Germany}}}
\affil[2]{\orgdiv{Institute of Applied Optics and Biophysics}, \orgname{Friedrich Schiller University Jena}, \orgaddress{\street{Philosophenweg 7}, \city{Jena}, \postcode{07743}, \country{Germany}}}

\abstract{\unboldmath
Measuring the phase of light is fundamental to optical imaging, sensing, and signal processing applications. Conventional optical phase measurements rely on multipath configurations, bulky interferometric setups, and computationally intensive data pipelines, limiting scalability, robustness, and practicality. We introduce a technique that allows for reference-free in-line phase retrieval of abrupt phase transitions in optical pulses directly from spectral measurements. Theory, simulations, and experiments concurrently explain the effect as a result of a Kerr-mediated interference between a projected linear wave component and a parametrically amplified residual of the phase-altered pulse. Utilizing this phenomenon, we demonstrate algorithm-free phase measurements of up to $\pi/385$\,rad sensitivity and shot-to-shot signal prominence at 13\,dB above noise at 80 MHz rates and 50\,pJ pulse energies. This approach offers new paths toward the use of femtosecond pulses as broadband data carriers for optical communications, information processing, and direct high-throughput phase imaging.
}


\maketitle

\renewcommand\linenumberfont{\normalfont\fontsize{10}{12}\selectfont}%

\pagebreak

\section{Introduction}\label{sec1}

Phase has become an indispensable resource of information across a wide branch of optical technologies. In bioimaging and metrology, phase provides label-free contrast for transparent specimens \cite{biomedicineQPI_ref1,tomography_ref1,production_ref1}. In photonics, the spectral phase of ultrafast pulses is a versatile control parameter enabling ultrafast pulse shaping \cite{pulseshaping_ref1,fsSLM_ref1} for advanced applications, e.g., in laser engineering \cite{Mao2021, Zhang2025}, nonlinear system control \cite{Wright2015}, spectroscopy and molecular control \cite{Weiner1990, Solli2008}, as well as optical information processing \cite{MZI_ref1,Mahjoubfar2017,Huang2019,Azana2025}. The ability to accurately read spectral phase becomes as critical as the ability to shape it.
As coherent ultrafast applications grow more complex and application demands more stringent, the ability to accurately read spectral phase becomes as critical as the ability to shape it.

Yet, retrieving phase from intensity-only measurements remains a fundamental challenge and requires prior assumptions, reference beams or often ambigious algorithmic inference. Interferometric and holographic approaches \cite{holographyQPI_ref2} provide the gold-standard as they offer high precision. Yet, free-space arrangements are rather bulky and environmentally sensitive. Fiber-based \cite{SMFinterferometer_ref2,UMZI_ref2,fiber_roztocki2021arbitrary} and on-chip \cite{onchipMZI_ref2} interferometers are more compact but require active stabilization of their multi-path arrangements. Self-referencing techniques, such as FROG \cite{FROG_ref2}, SPIDER \cite{SPIDER_ref2}, and SRSI \cite{SRSI_ref2}, have become workhorses of ultrafast characterization, yet rely on complex free-space alignment, require nano- to microjoule pulse energies that typically necessitate amplification, and carry the intrinsic ambiguities of iterative phase retrieval. Even though deep learning approaches have recently dramatically accelerated convergence of classical iterative algorithms \cite{gerchberg1972_ref2,retrivalALG_ref2,DL_ref2,realPYT_ref2,fourierPYT_ref2}, they introduce concerns around generalizability and interpretability beyond the data used for their training. Together, these constraints motivate the search for a simpler alternative: one that operates directly on intensity, requires no reference or iterative algorithm, and functions at oscillator-level pulse energies.

Here we demonstrate that weakly nonlinear Kerr propagation in standard optical fiber provides precisely this capability. We demonstrate a reference- and algorithm-free technique for spectrally distributed phase recovery using weak nonlinear phase modulation in fiber. We reveal, for the first time, the origin of this new phenomenon through theory, single- and coupled-mode simulation, and experiment, and show that weak nonlinear Kerr effects can transform spectral phase shifts directly into intensity peaks.  We establish the physical origin of this phenomenon and identify wide operating ranges across pulse width, peak power, and phase magnitude. Using off-the-shelf fiber components in a fully integrated setup, we demonstrate the technique in two spectrometric configurations: grating-based measurements yield a phase sensitivity $\pi/385$\,rad, while shot-to-shot measurements at MHz rates and picojoule energy-level achieve above 13\,dB SNR without any post-processing. Overall, this work introduces a new application field for nonlinear fibers as a practical platform for direct phase sensing, imaging, and optical information processing.

\section{Results}\label{sec2}
\subsection{Self-referenced Kerr-induced Interferometry}\label{sec21}

\begin{figure}
\centering
\includegraphics[width=\textwidth]{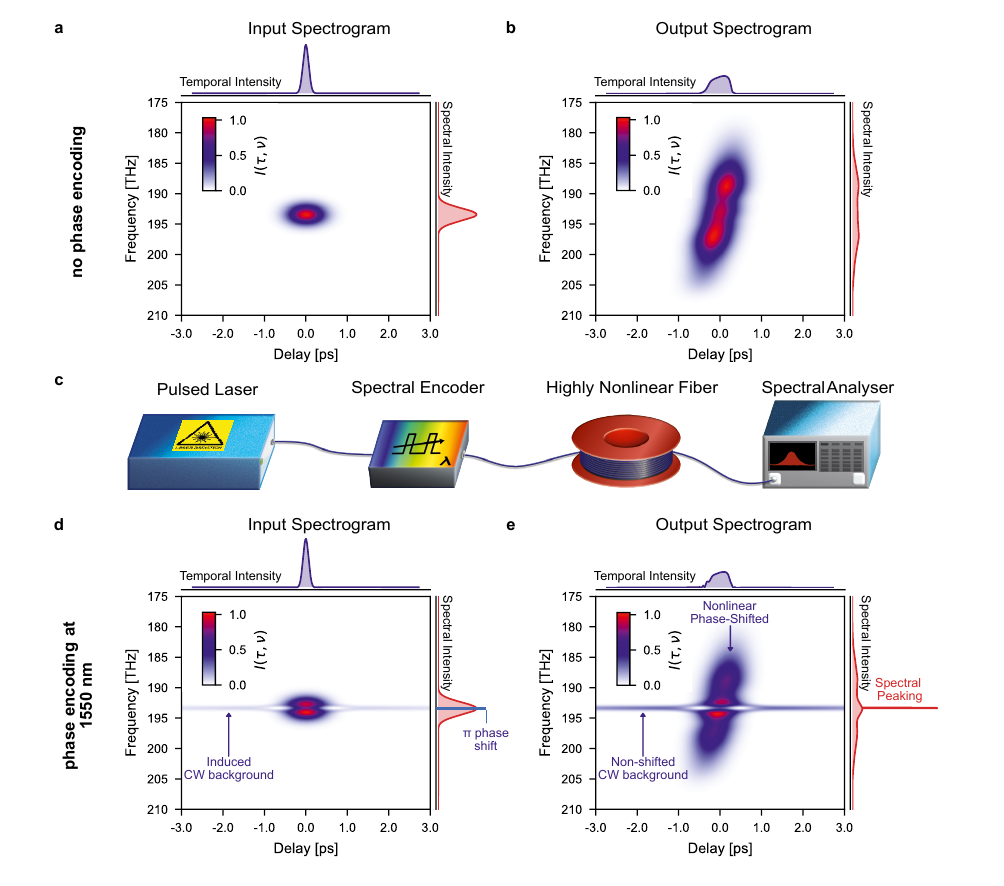}
\caption{Experimental setup and spectrograms illustrating the nonlinear phase modulation that leads to spectral peak formation. (a, b) Spectrogram at (a) input and (b) output of 5\,m highly nonlinear fiber (HNLF), showing the temporal and spectral intensity profiles of a Fourier-limited pulse. (c) Schematic of the experimental setup, consisting of a femtosecond pulsed laser centered at 1550 nm, a spectral phase encoder, an HNLF, and a spectral analyzer for readout. (d, e) Input and output spectrogram for a 1\,nm-wide $\pi$-phase shift centered at 1550nm. A continuous-wave (CW) background appears in the temporal domain due to the imposed spectral phase modulation and persists at the output, resulting in a pronounced spectral peak at the modulation frequency. Note that the higher contrast of the CW wave in (e) occurs due to a renormalisation of the spectrogram plot. The intensity scales on the left axis of panels (b) and (e) are equal.}\label{Figure 1}
\end{figure}

The propagation of intense electromagnetic pulses through optical Kerr media is commonly accompanied by nonlinear effects, such as self-phase modulation, leading to characteristic modulations in the measured intensity spectrum (Fig. \ref{Figure 1}a,b). Those spectral modulation, while sensitive to pulse power, shape, and phase (e.g., chirp) usually do not directly reveal phase information of the incident pulses.

\begin{figure}
\centering
\includegraphics[width=\textwidth]{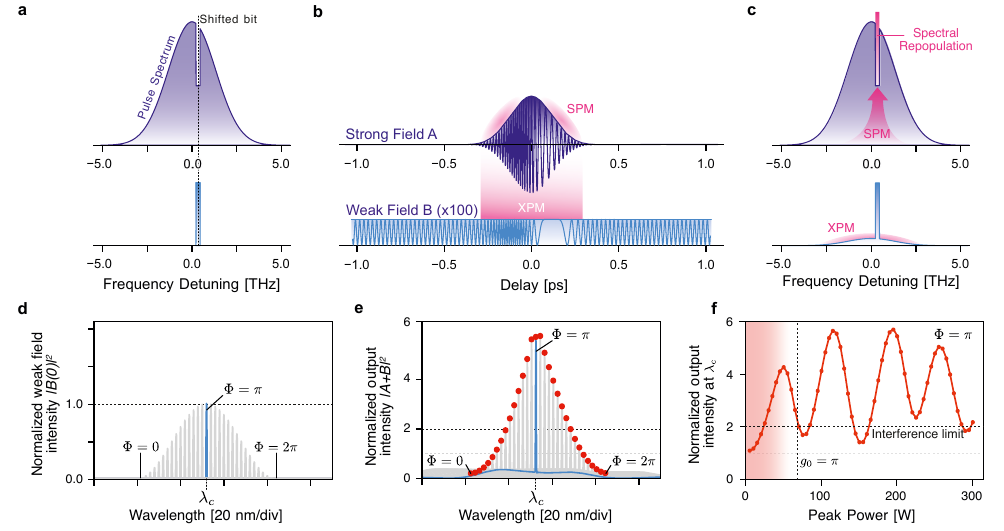}
\caption{Self-seeded parametric interference in coupled-mode theory. (a) Model input configuration illustrating the field decomposition into a spectrally notched strong field spectrum (top, blue) and phase-shifted weak field detuned by $\omega_c$ from the center frequency (bottom, purple). (b) Time-domain representation of the strong pulse field A (top) and weak CW field B (bottom) showing SPM and XPM phase modulations acquired during nonlinear propagation. (c) Output spectra after Kerr propagation, illustrating the fill-up of the spectral notch due to SPM-broadening (top) and XPM-modulated spectral component (bottom) at the probe frequency. Note that the initial relative phase at $\omega_c$ between the strong and the weak field has also changed (not shown here). (d-e) Input spectrum of the weak field (d, blue) and output spectrum of both fields combined (e, blue) as a function of wavelength for varying phase encoding $\Phi$ from 0 to 2$\pi$ according to coupled-mode theory, showing characteristic spectral peaks at probe wavelength $\lambda_c$. Spectra in grey were artificially shifted by 30\,nm/rad as a function of $\Phi$ to illustrate the change in peak intensity (red markers). All spectra are normalized to the weak field amplitude at input for $\Phi=\pi$\,rad. (f) Peak spectral intensity at the probe wavelength for $\Phi = \pi$\,rad versus input peak power $P_0$, demonstrating nonlinear phase-sensitive amplification with values exceeding the typical constructive interference limit of factor 2. The shaded area indicates the weakly nonlinear regime.}
\label{Figure 2new}
\end{figure}

A nonlinear medium, like a highly nonlinear optical fiber, can reveal discontinuous spectral phase modifications through intra-pulse interference between linear and nonlinear wave contributions. To formalize this phenomenon, we begin by expressing the pulse as a complex field in the spectral domain and introduce a narrowband flat-top phase filter (referred to as \textit{bit}), defined as the combination of two Heaviside phase steps (see Appendix A). The temporal impulse response $h(t)$ of such a phase filter with shift $\Phi$, we found to be
    \begin{equation}
    h(t) \approx \delta(t) + (e^{i\Phi} - 1) \frac{\Delta\omega}{2\pi} \mathrm{sinc}\left(\tfrac12 \Delta\omega t \right) e^{i\omega_c t} \ ,
    \end{equation}
where $\omega_c = \tfrac{1}{2}(\omega_{s,1} + \omega_{s,2})$ is the central wavelength and $\Delta\omega = \omega_{s,2} - \omega_{s,1}$ is the bandwidth of the bit. If we consider narrow bit widths $\Delta\nu$ below 100\,GHz, the temporal span of the sinc, defined as the coherence time $T_c=\tfrac{1}{\Delta\nu} \geq 10$\,ps, significantly exceeds the pulse width of sub-picosecond pulses. The sinc can hence be approximated to a first-order for arguments near zero with
     \begin{equation}
    h(t) \approx \delta(t) + (e^{i\Phi} - 1) \frac{\Delta\omega}{2\pi} e^{i\omega_c t} \ .
    \end{equation}
Convolving this response with the temporal field $A(t)$ yields a continuous-wave (CW) background (Fig.\ref{Figure 1}d) that overlaps with the remaining temporal components of the pulse: 
    \begin{equation}
    f(t) = \underbrace{A(t)}_{\text{Pulse}} + \underbrace{(e^{i\Phi} - 1) \frac{\Delta\omega}{2\pi} a(\omega_c) e^{i\omega_c t}}_{\text{CW component at } \omega_c} \ . 
    \end{equation}
Hence, an abrupt change to the spectral phase causes, in first approximation, a modified pulse and a CW contribution at the spectral location of the introduced bit. Notably, the intensity of the weak field, and thus the energy redistributed from the pulse, is phase dependent and follows a $\sin^2\tfrac{\Phi}{2}$ behaviour, as shown in Fig. \ref{Figure 2new}d.

Propagating this self-referenced composite field through a highly nonlinear fiber or a waveguide causes a highly non-trivial dynamics: if perturbations remain small, the strong field undergoes self-phase modulation (SPM) and additionally modulates the phase of the weak field due to cross-phase modulation (XPM) (see Fig. \ref{Figure 2new}b). The interference of both phase-shifted contributions $|A(z)+B(z)e^{i\omega_ct}|^2$ reveals the phase in the measured intensity as a striking spectral feature, visible in Fig. \ref{Figure 1}e. 

Within the weakly nonlinear limit, the relative intensity upon interference at $\omega_c$ can be found to be
\begin{equation}\label{eq:R_def}
  R \;\equiv\; \frac{|a'(\omega_c)|^2}{|a(\omega_c)|^2}
  \;=\; |\alpha + \beta|^2
  \;=\; |\alpha|^2 + 2\,\mathrm{Re}[\alpha^*\beta] + |\beta|^2\,,
\end{equation}
with $|a(\omega_c)|^2$ and $|a'(\omega_c)|^2$ being the spectral intensity at the input and output of the nonlinear waveguide, respectively. The linear-XPM coefficient $\alpha$ and the SPM ratio $\beta$ are defined as
\begin{equation}\label{eq:alpha_beta}
  \alpha = e^{i\Phi} + \frac{i(e^{i\Phi}-1)\,\Delta\omega\,\Gamma L\,E}{2\pi^2}\,,
  \qquad
  \beta = \frac{i\Gamma L}{2\pi}\,\frac{[a*S](\omega_c)}{a(\omega_c)}\,,
\end{equation}
where $\Gamma$ is the nonlinear parameter of the waveguide, $L$ is the waveguide length, $E = \int|a(\omega)|^2\,d\omega$ is the pulse energy and $S(u) =
\tfrac{1}{2\pi}\int a(\omega)a^*(\omega-u)\,d\omega$ is the spectral auto-correlation
(see Appendix B for the full derivation).

To illustrate the dependence on phase and peak power, we may assume a transform-limited Gaussian pulse probed at its carrier frequency ($\omega_c = \omega_0$). Doing so, $\beta$ evaluates to a purely imaginary number $\beta = ig_0/\!\sqrt{3}$, where $g_0 = \Gamma L P_0$ is the nonlinear gain with peak power $P_0$ of the optical pulse. Substituting into Eq.~\eqref{eq:R_def} and retaining only the leading XPM correction yields a closed-form result
\begin{equation}\label{eq:R_Gauss}
  R(\Phi,\,g_0) = 1 + \frac{2\,g_0}{\sqrt{3}}\,\sin\Phi + \frac{g_0^2}{3}\,.
\end{equation}
Two key features emerge directly from Eq.~\eqref{eq:R_Gauss}: First, the phase sensitivity is \emph{sinusoidal}, i.e., the interference term $(2g_0/\!\sqrt{3})\sin\Phi$ is maximised at $\Phi = \pi/2$ and vanishes at $\Phi = 0$ and $\pi$, where the phase-rotated input field is orthogonal to the purely imaginary SPM contribution and no coherent beating occurs (Fig.~\ref{Figure 2new}e). Second, at fixed $\Phi$, $R$ grows linearly with power in the weak-gain regime ($g_0 \ll 1$) before the quadratic SPM floor $g_0^2/3$ takes over, setting a practical operating window for analog phase read-out (Fig.~\ref{Figure 2new}f).

While illustrative, Eq.~\eqref{eq:R_Gauss} does not reveal the full dynamics of the phenomenon. In particular, experimental observations show spectral peaks with much greater prominence than expected from interference alone. The perturbative treatment requires $g_0 < \sqrt{3}$, beyond which the weakly nonlinear approximation breaks down. Moreover, the input spectrum $a(\omega)$ is assumed unperturbed, i.e., unmodified by the phase encoding, whereas in reality the energy transferred into the CW mode notches the pulse spectrum at $\omega_c$ in a $\Phi$-dependent manner (cf. Fig.~\ref{Figure 2new}).

Coupled mode theory allows us to address both limitations. Solving the coupled equations for the strong field $A$ and the weak field $B$ (see Methods) reveals several key observations: First, SPM leads predominantly to energy transfer to the spectral notch, an effect that has been observed earlier for amplitude-filtered pulses by Pr\"akelt \emph{et al.} \cite{2005filling}. Second, this spectral repopulation, as illustrated in \ref{Figure 2new}c, can exceed the original spectral intensity significantly, leading to a considerable spectral amplification above the typical factor-of-two bound of a coherent interference between equally split waves when combined with the weak field, as shown in Fig. \ref{Figure 2new}e. Lastly, the spectral intensity peak indeed oscillates for peak powers, and hence nonlinear gain, beyond the weakly nonlinear limit, as shown in Fig. \ref{Figure 2new}f, which is a result of the nonlinear phase rotation.

We refer to the entirety of the nonlinear interference dynamics as Kerr-induced in-line interferometry (KI3) to avoid confusion with the nonlinear interference in $\chi^{(2)}$-nonlinear crystals \cite{Graefe}. 
The key characteristic of Kerr-induced interferometry features a distinct, power-dependent intensity peak after nonlinear propagation at spectral locations where the femtosecond pulse undergoes an abrupt phase change. 

\subsection{Single-Bit Analysis}\label{sec22}

To experimentally verify the nonlinear phase-to-intensity mapping (Fig. \ref{Figure 1}c), a narrowband phase bit was introduced on the laser spectra using a programmable spectral phase modulator (WaveShaper). The modulated pulse is then propagated through 5\,m of normal dispersive, highly nonlinear fiber (HNLF). The nonlinear interference occurs during propagation and the resulting phase-to-intensity mapping can be readout using an optical spectral analyzer (OSA).

\begin{figure}
\centering
\includegraphics[width=0.5\textwidth]{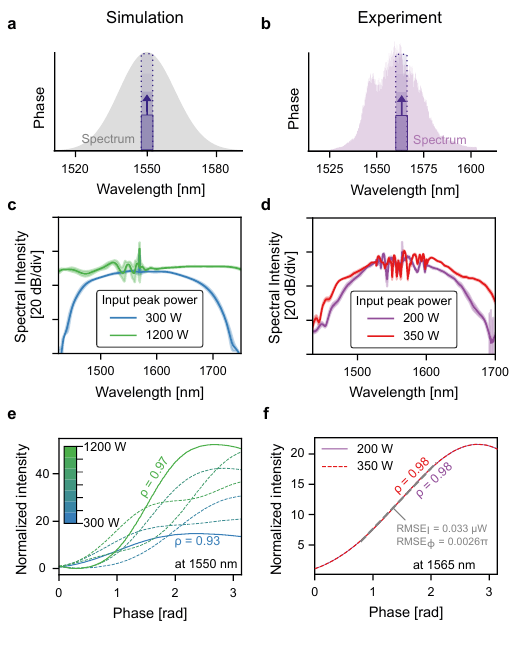}
\caption{Comparison of narrowband spectral phase encoding ("bit") in full-field simulation and experiment. (a, b) Schematic illustration of narrowband flat-top phase encoding of a single analog bit, shown for (a) simulation and (b) experiment. The bit (1nm wide) is not to scale relative to the laser spectrum. The arrow indicates that phase of the bit is tuned from 0-$\pi$. (c, d) Average output spectra with standard deviation (shaded) for low and high input peak powers. (c) Simulations based on the nonlinear Schödinger equation for 300\,W and 1200\,W peak input power. (d) Experimental spectra for 200\,W and 350\,W. Distinct spectral peaks are distinguishable only at low power in the experiment. (e, f) Correlation between output spectral intensity, normalized to spectral intensity at phase $\phi = 0$, and input phase at the encoding wavelength. (e) Simulation results show overall high Pearson correlation coefficients ($\rho$) for 300\,W (solid blue) and 1200\,W (solid green), with intermediate powers (dotted lines, power according to colorbar) in between. (f) Experimental curves exhibit a similar trend, with a Pearson correlation of 0.98.}
\label{Figure 2}
\end{figure}

Since fiber laser setups are commonly limited in peak power, we studied the power dependency of the effect numerically. 
Near-linear phase-to-intensity mapping persists across a broad power range. Our simulations for peak power levels from 300 to 1200\,W show the intensity-mapped phase information remains in the spectral position, remarkably even when the spectrum significantly broadens due to nonlinear frequency generation (Fig. \ref{Figure 2}c). At this spectral position, we observe a close to linear dependency of the output intensity to the input phase magnitude. The Pearson correlation, defined by $\rho =cov(I,\phi)/(\sigma_I \sigma_\phi)$, where $I$ represents intensity and $\phi$ represents phase, is close to 1 for all power levels supporting the linear trend (Fig. \ref{Figure 2}e).  

Experimental measurements with 200\,fs pulses at peak powers of 200 and 350\,W confirm the linear phase-to-intensity mapping, despite considerable spectral distortions at higher powers. Measurements were performed at 200 and 350\,W peak power and 200\,fs pulse duration. 
Pulse and fiber imperfections, as well as additional nonlinearity arising from optical amplification before encoding, lead to differences between measured and simulated spectra. Nonetheless, the general linear trend between phase and peak amplitude at the bit‘s spectral location remains for all tested settings in simulation and experiment. Even in cases where the peak prominence severely declines in measured data (e.g., see Fig. \ref{Figure 2}d or Fig C5 in Appendix C), the spectral output at bit position shows strictly linear behavior, featuring a high Pearson correlation of 0.98 (Fig.\ref{Figure 2}f). 

Remarkably, we obtain an extraordinary phase sensitivity of $\pi/385$\,rad within the linear range of $\pi/2$ to $2\pi/3$\,rad from the root-mean-square error of the linear fit (see Appendix C2). This is well within the current state-of-the-art in phase retrieval techniques, where recent techniques, such as Zernike wavefront imaging microscopy, have reported sensitivities of up to $\pi/655$\,rad \cite{Gentner2024}.

The effect is well scalable to multiple frequencies. Repeating the experiment with two individual phase bits shows that nearby bits remain spectrally independent, with their spectral responses acting on in-band phase variations but not on changes in neighboring bands. Both simulations and experiments confirm that the spectral positions of the two bits are preserved (see Appendix D).

\subsection{Distributed Phase Recovery}\label{sec23}

\begin{figure}
\centering
\includegraphics[width=\textwidth]{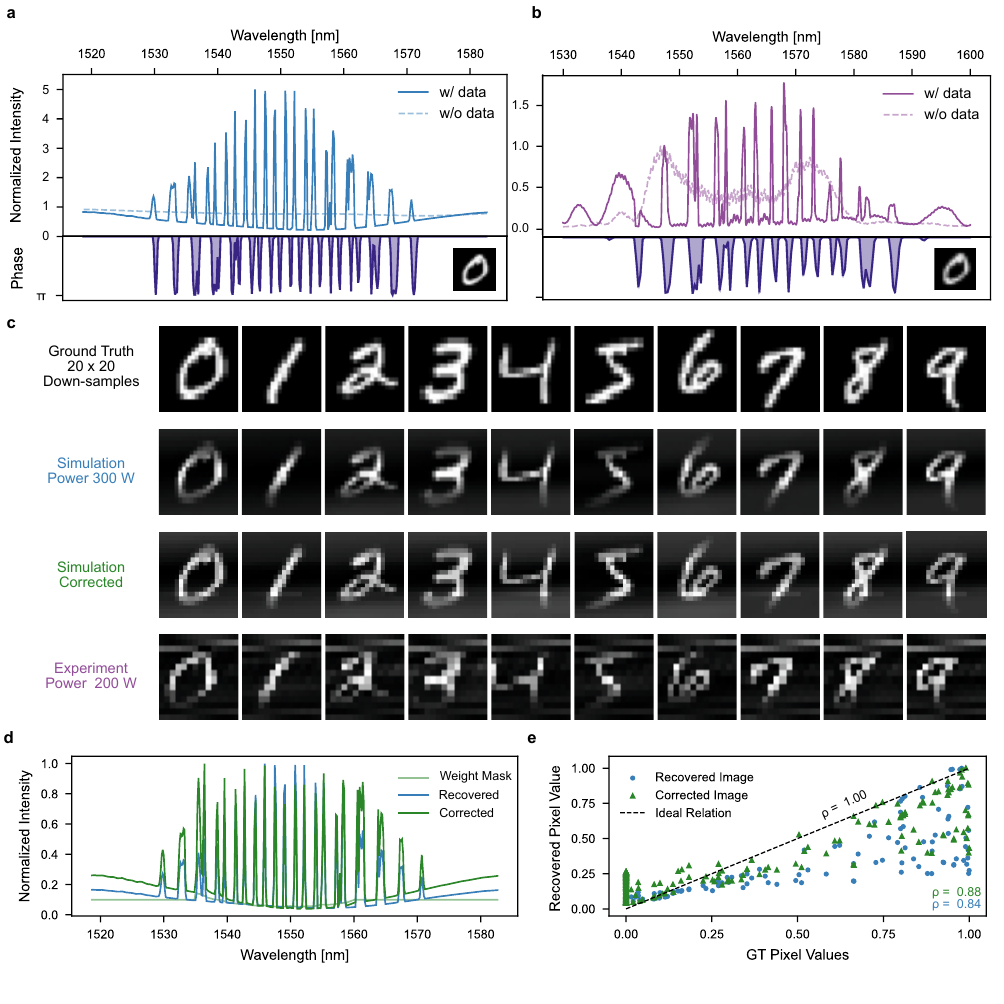}
\caption{Comparison of simulated and experimental spectral encoding and phase recovery using MNIST digits. (a) Simulation results showing the input phase mask (bottom, inverted y-axis) and corresponding normalized output spectrum (top). The inset shows the 20×20 ground-truth (GT) image used for simulation. (b) Experimental results showing the encoded phase mask and measured spectrum. The inset displays the corresponding 14×14 ground-truth image used in the experiment. (c) Recovered images from top to bottom: The first row shows the 20×20 ground-truth digits 0–9 from the MNIST dataset (used for simulation), the second row shows simulation-based recovery at input peakpower of 300W, the third row shows the recovered images spectral weighting–based correction of spectral distortions, and the fourth row shows experimental recovery from raw spectral data at input peak power ~200\,W and pulse width of 200\,fs. (d) Normalized spectra for a representative sample of digit "0", showing the measured spectrum used for the recovered image (blue), the envelope-based weighting function obtained from the single wavelength shifted phase bit (light green), and the corrected spectrum after applying this weighting (green). (e) Pixel-value correlations of ground truth vs. the recovered image (blue) and GT vs. the corrected image (green) are shown relative to the ideal linear relation (black dashed line). }
\label{Figure 3+4}
\end{figure}

The in-fiber phase-to-intensity projection is ideally suited to recovering broadband spectral phase information when stored in an optical pulse. To demonstrate the capability of Kerr-induced in-line interferometry to image spectrally distributed phase information, we illustrate the principle using MNIST grayscale images of handwritten digits (0–9) \cite{mnist} as input. The original 28$\times$28 images were resized to 20$\times$20 for simulations and 14$\times$14 for experiments (limited by the encoder and readout resolutions), then flattened. The flattened arrays with normalized entries from 0 to 1 were multiplied by a maximum phase shift of $\pi$ rad (Fig. \ref{Figure 3+4}a,b bottom). These phase masks were applied to 200 wavelength bins of the input spectra, centered at 1550 nm in simulation and between 1525-1600 nm in experiment, and were then propagated through the nonlinear fiber. At readout, the distributed phase bits occur as spectral intensity peaks, spectrally well aligned with the input mask (Fig. \ref{Figure 3+4}a,b, top). 

The input phase information can be directly extracted from the measured spectra at the encoded wavelengths. The spectra are cropped to the spectral domain of the input phase mask. The cropped one-dimensional spectral intensity data are then mapped back into a two-dimensional grid of the original input format (20x20 in simulation, and 14x14 in experiment). This “reshaping” assumes a linear correspondence between spectral bins and pixel indices. Since the number of recorded spectral samples does not exactly match the number of encoded pixels, linear interpolation is applied to resample the data onto the two-dimensional grid. Recovered images match the ground truth well for qualitative inspection (Fig. \ref{Figure 3+4}c). This reconstruction demonstrates that distributed spectral phase information can be retrieved directly from intensity-only measurements.

Signal distortions may occur from two dominant sources: (a) dependence of output spectral intensity on the input spectral profile, leading to a decline of the peak prominence at the spectral edges (less prominent in experiment due to broader and flatter laser bandwidth), and (b) experimental imperfections, such as non-symmetric or chirped input pulses, which distort the spectral baseline. Straightforward correction techniques, such as normalizing readouts to the spectral baseline or the system response, can reduce those artifacts (see Methods). Importantly, all aberrations could, in principle, also be compensated optically through ideal pulse compression and spectral flattening of the input.

\begin{figure}
\centering
\includegraphics[width=\textwidth]{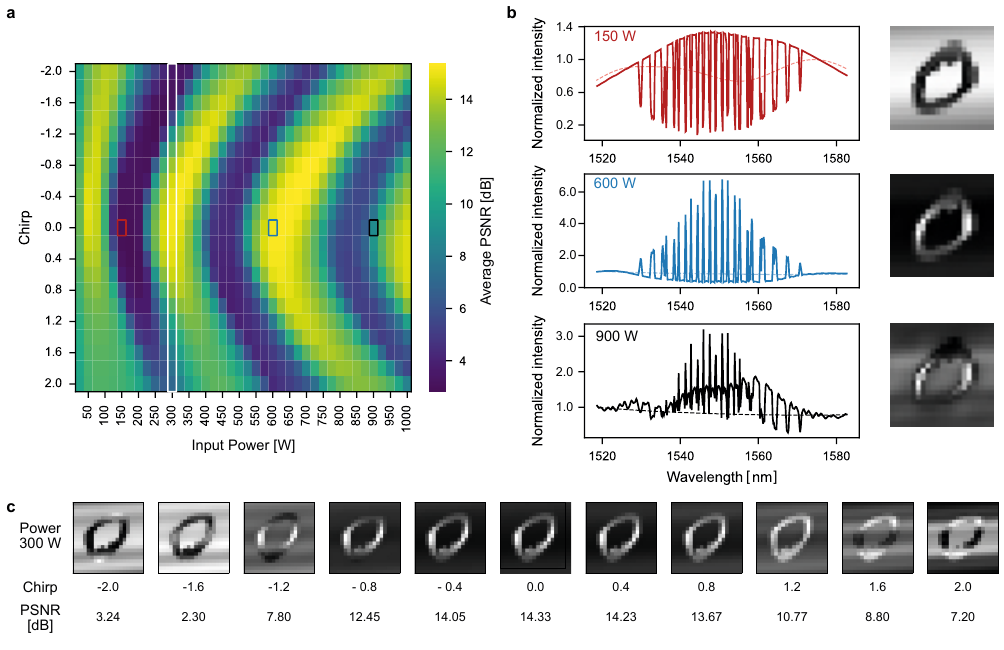}
\caption{Dependency of phase recovery fidelity on initial peak power and chirp. (a) Heatmap of average peak signal-to-noise ratio (PSNR) as a function of input peak power and applied spectral chirp (–2 to +2) in simulation. Each point represents the mean PSNR over 10 reconstructed digits (0–9) compared with their respective ground truths. (b) Influence of peak power on image recovery at zero chirp. Left: normalized output spectra without data (dotted line) and with data for digit "0" (bold line)  at 150 W, 600 W, and 900 W input peak powers. Right: corresponding recovered digit images showing change in recovery fidelity with power. (c) Effect of chirp on phase recovery at fixed input power (300 W). Recovered images of digit “0” are shown for chirp values from -2 to +2, illustrating systematic degradation and phase distortion away from zero chirp at this power.}
\label{Figure 5}
\end{figure}

Because of the nonlinear nature of this phase mapping technique, input peak power and pulse width become important control parameters to choose with respect to the available waveguide length \cite{ramanPeak_discussion}. In simulation and experiment, we adjusted the pulse width by controlling the spectral chirp to maintain the spectral bandwidth and hence the information capacity of the pulses. We quantify recovery fidelity for each setting using the peak signal to noise ratio (PSNR), a measure of the quality of an image compared to its original (see Methods), averaged over ten digits for various pulse width and power combinations (Fig. \ref{Figure 5}a). The PSNR measures the similarity between our recovered and original images. The modelled PSNR map in Fig. \ref{Figure 5}a unambiguously reveals the characteristics of the nonlinearity-induced interference: distinct operational regions oscillate between optimal and suboptimal performance as a function of pulse power and chirp. 

Representative spectra and corresponding digit recoveries for three different powers at zero chirp are shown in Fig. \ref{Figure 5}b. At 600 W, the recovered image closely matches the ground truth with minimal background. At 900 W, the digit is faint and distorted by the nonlinear background. At 150 W, contrast inversion appears due to destructive interference. These results demonstrate that the phase can be recovered directly if the system is operated at an appropriate power for a given pulse width and fiber length. To further illustrate the role of input conditions, Fig. \ref{Figure 5}c shows the effect of varying input chirp at fixed power (300 W). Increasing positive or negative chirp systematically degrades recovery fidelity in this power domain. Outside the optimal operating range, contrast diminishes and distortions emerge due to the nonlinear background, leading to contrast inversion or filtering effects.

\subsection{Single-shot Phase Recovery}\label{sec24}

\begin{figure}
\centering
\includegraphics[width=\textwidth]{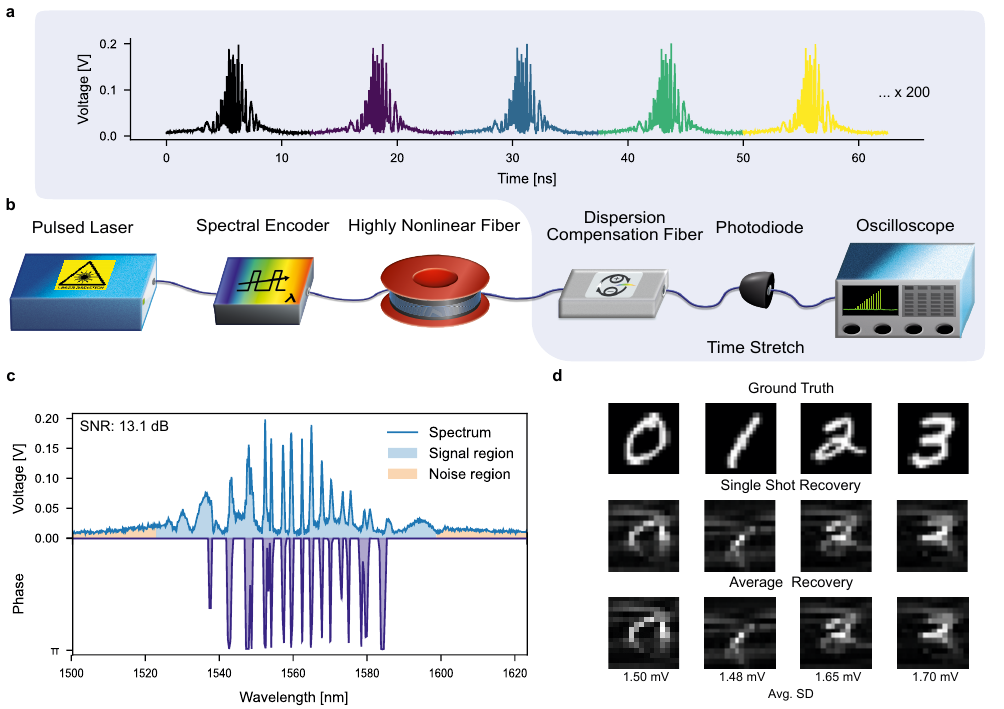}
\caption{Single-shot spectral acquisitions and recoveries. (a) Five representative single-shot measurements from a total of 200, obtained via dispersive spectrum-to-time acquisition of spectrally encoded pulses. (b) Modified experimental setup where the spectral analyzer is replaced with a dispersive Fourier transform unit comprising dispersion compensation fiber (DCF) to stretch the spectrum over time, an ultrafast photodiode, and a real-time oscilloscope for intensity readout. (c) First of the 200 time-stretch shots, mapped to wavelength (top), and the applied phase mask of digit "0" (bottom). (d) Ground truth (top),  recovered images from the single time-stretch shot (middle), and averaged recovery of digits over ~200 shots, with their respective mean standard deviation (SD) across all pixels listed below.}
\label{Figure 6}
\end{figure}

Finally, we demonstrate a proof-of-concept experiment for reference-free pulse-to-pulse phase recovery. To capture spectral phase information at pulse rates, we employ dispersive Fourier transform (DFT) \cite{DFT}, a versatile technique that converts spectral content into measurable temporal traces. In this method, a dispersion compensation fiber (70\,ps/nm/km) introduces wavelength-dependent delays that map the spectral profile of each pulse into the time domain. The time-stretched pulses are then detected with a 33\,GHz high-speed photodiode in combination with a 110\,GHz real-time oscilloscope with 10\,bit analog channel precision (Fig. \ref{Figure 6}b). An example of recorded time traces is shown in Fig. \ref{Figure 6}a.

We performed measurements at an 80 MHz repetition rate using 200 fs input pulses and 200 W pump power. From the recorded time traces, we successfully retrieved distributed spectral phase information and recovered MNIST images on a pulse-to-pulse basis. A single-shot recovery yields a signal-to-noise ratio (SNR) of 13 dB (Fig. \ref{Figure 6}c), establishing the viability of pulse-to-pulse phase extraction. Averaging over 200 independent shots yields a mean pixel-wise standard deviation of 1.51\,mV, which is 2.47$\times$ lower than the noise floor of the spectral background (3.7\,mV). Representative recovered images are shown in Fig. \ref{Figure 6}d.

The misalignments in the DFT wavelength mapping (Fig. \ref{Figure 6}c) stem from inherent dispersion stretching. Different wavelengths experience different group delays; thus, shorter wavelengths may compress into narrow time windows, whereas longer wavelengths spread across broader time intervals, creating an inherently nonlinear mapping between time and wavelength. Another possible contribution is the peak-finder mapping misalignments arising from numerical quantization and localization errors (see Appendix C4). 


Our demonstration achieves repetition rate and fidelity comparable to previous works on single-shot phase reconstruction techniques \cite{80Mhz,46Mhz, singleshot2011, singleshot_noninterferometric2013, singleshot2020}. The recovery in our setup is limited by detector sensitivity and noise, as well as the limited bit depth of the oscilloscope ADC, which defines the effective baseline of each trace. Additional losses from fiber connectors and propagation further degrade SNR. Improvements in detector bandwidth, ADC resolution, and optical coupling efficiency are therefore expected to significantly enhance performance. 

This highlights the potential of Kerr-induced in-line interferometry for extending ultrafast phase characterization beyond existing limits. Overall, these results demonstrate the feasibility of reference-free, high-rate pulse-to-pulse phase sensing, with direct implications for the real-time characterization of optical devices and ultrafast communication systems.

\section{Discussion}\label{sec3}
Our theoretical framework provides the first general understanding of the effect. The interplay between self-phase modulation of the strong field, which leads to parametric amplification at $\omega_c$, and cross-phase modulation of the weak field may cause constructive interference between the two fields. The peak intensities of this Kerr-induced in-line interference can significantly exceed the factor of two of typical constructive interference between two equally split waves. The Kerr-induced spectral amplification manifests as oscillations as a function of power, as observed in simulations based on coupled-mode theory and the broadband nonlinear Schrödinger equation. The model does not impose limits on the range of observation with respect to pulse width, peak power, fiber length, or fiber dispersion (see Supplementary sec. B3). Such generality indicates that the effect is not bound to critical operating conditions but is instead a fundamental consequence of abrupt phase modification and of the nonlinear–linear interference mechanism. 

However, the signal prominence is indeed dependent on the configuration of the nonlinear system. Simulations help to identify operational domains and clarify where the effect becomes most distinct from the SPM-broadened spectral background. In our study, the peaking effect remained visible even up to the tested domain limits of pulse chirp and peak power. Notably, particularly in the highly nonlinear regime, SPM not only induces parametric amplification at the notch but also redistributes energy from the vicinity of the peak signal over a wide bandwidth, thereby exposing the phase-modified spectral peak. Thus, exceptional peak prominence could be observed of factors of 10 or more above the spectral background. On the downside, nonlinear spectral modulations become dominant at peak powers $>500$\,W (chirp-free), introducing a nonlinear background and degrading the signal fidelity. 

The persistence of the technique across such a broad parameter space makes it difficult to suppress this effect. This robustness may have implications for optical signal processing and information processing tasks in which femtosecond- to picosecond-pulse pulses play a pivotal role. For example, evidence from fiber-based extreme learning machines indicates that the spectral window used to encode information can preserve high information content even after significant spectral broadening via supercontinuum generation \cite{Brunner,Saeed}. In modern telecom systems, ever-increasing modulation speeds (100 GHz and above) generate pulses with durations $\leq 10$\,ps or shorter and considerable peak power. Given an ideal instantaneous phase-modulation response and high amplification levels, the effect could introduce a source for intra-channel signal deterioration. Furthermore, narrow phase or amplitude changes (e.g., gas absorption lines or filter windows) may project phase information into amplitude after amplification and propagation, even in weakly nonlinear media, with unforeseen consequences for signal purity in laser engineering and sensing \cite{modelock}.

\section{Conclusion}\label{sec4}
In summary, our study establishes a universal theoretical framework for utilizing a new nonlinear phenomenon in Kerr media and verifies its validity with experiments on a normal-dispersion fiber. The combined theoretical and experimental results demonstrate that the effect persists across a broad range of system parameters, making it attractive for low-power phase retrieval applications. Our framework explains and generalizes earlier empirical observations of related spectral effects reported in a range of experimental studies\cite{westbrook2004SC, 2005filling, yeom2007tunable, ramanPeak_discussion, okazaki2022mode, li2026linear}. Here, we showcased this method for ultrafast phase sensing and imaging, achieving pulse-to-pulse readout at tens of MHz rates and picojoule pulse energies using dispersive Fourier transform. This highlights its potential for reference-free and algorithm-free characterization of optical systems at speeds exceeding those of conventional interferometric techniques. Only a single calibration step is required to obtain accurate quantization, assigning an absolute phase value to the measured intensity, with a sensitivity of $1/385 \pi$\,rad, estimated from the MSE fit error.

Overall, the universality of the effect has implications for optics and photonics, in particular laser engineering, high-throughput phase imaging and reconstruction, and optical signal processing. This work also highlights the potential of femtosecond pulses as information carriers for future communications and computing, as well as sensitive probes of complex light–matter interactions. 

\section{Methods}\label{Methods}

\subsection*{Experimental setup}\label{Methods: setup}
Experiments were conducted on a vibration-isolated optical table under stable laboratory conditions. A mode-locked fiber laser (Toptica DFC Core) with 80\,MHz repetition rate, 1556\,nm central wavelength, 830\,fs pulse duration, and 13\,mW average power, served as the source. The output was coupled into 1m long polarization-maintaining fiber, ensuring stable polarization.

A programmable optical WaveShaper (Coherent Waveshaper 1000A) was used to apply user-defined phase modulation. The WaveShaper device, which consists of a grating and a liquid crystal on silicon (LCoS) spatial light modulator, enabled modulation of phase across $>300$ spectral channels of the pulse spectrum, and hence is used in our setup to a) compresses the 830 fs broad pulse of the laser to a near-transform-limited ($\sim$200 fs) using genetic algoritm based optimization \cite{GA} and b) implement spectrally localized phase shifts. 

The modulated light was then propagated through a normally dispersive highly nonlinear fiber (Thorlabs HN1550, length 5m, dispersion -1 ± 1 ps/nm/km). The average power was monitored with a calibrated power meter (Thorlabs PM100A). The temporal pulse width and profile was measured using an autocorrelator (APE, PulseCheck NX50). The spectra were recorded with an optical spectrum analyzer (Yokogawa AQ6375E, resolution 0.02 nm, wavelength range: 1000-2500nm).

For dispersive Fourier transform (DFT) measurements, the pulses output from the HNLF were passed through 800 m of dispersion-compensating fiber (M2 Optics, Inc.) with an overall dispersion of 70 ps/nm/km. The temporally stretched pulses were detected using a high-speed photodiode (Albis PMY30A-L, bandwidth 34.4 GHz) and digitized with a real-time oscilloscope (Keysight UXR1104B, bandwidth 110 GHz, sampling rate 256 GS/s, 10-bit ADC).

\subsection*{Numerical Simulation of Coupled-Wave Propagation}\label{sec:methods_CWE}

To validate the analytical predictions of Sec.~\ref{sec21}, propagation of the pump field $A(z,t)$ and probe field $B(z,t)$ has been modelled by two coupled nonlinear Schrödinger equations retaining chromatic dispersion, SPM, and XPM:
\begin{align}
\label{eq:CWE_A}
  \partial_z \tilde{A} &= i\beta(\omega)\,\tilde{A}
  - i\gamma\,\mathcal{F}\!\left\{\bigl(|A|^2 + 2|B|^2\bigr)A\right\},\\[4pt]
\label{eq:CWE_B}
  \partial_z \tilde{B} &= i\beta(\omega)\,\tilde{B}
  - i\gamma\,\mathcal{F}\!\left\{\bigl(2|A|^2 + |B|^2\bigr)B\right\},
\end{align}
where $\beta(\omega) = \tfrac{\beta_2}{2}\omega^2 + \tfrac{\beta_3}{6}\omega^3 +
\tfrac{\beta_4}{24}\omega^4$.  The pump is initialized as $A_0(t) = \sqrt{P_0}\exp(-t^2/2T_0^2)$
with $T_0 = 180$~fs ($T_\mathrm{FWHM} = 250$~fs), and the probe from the analytic
result of Appendix A.2 as
\begin{equation}
  B_0(t) = c\,e^{i\omega_c t}\,\mathrm{sinc}\!\left(\tfrac{\Delta\omega\,t}{2}\right),
  \qquad c = (e^{i\Phi}-1)\tfrac{\Delta\omega}{4\pi}\,a(\omega_c)\,.
\end{equation}
Equations~(\ref{eq:CWE_A})--(\ref{eq:CWE_B}) are integrated with a symmetric split-step Fourier method, using Runge-Kutta-4th-order integration for the nonlinear substep and the half-step operator $\exp[i\beta(\omega)\,\delta z/2]$ for dispersion, over $N_z = 1000$ steps across $L = 5$~m with the same $\Gamma$ and fiber dispersion coefficients as used in the broadband Schrödinger model below. The time grid comprised $N_t = 2^{14}$ points over a 50~ps window. Output spectra shown in Fig. \ref{Figure 2new}(d-f) are normalized to the peak spectral intensity of $B_0$ at
$\Phi = \pi$,
\begin{equation}
  G(z) \equiv \frac{|\hat{A}(\omega_c,z) + \hat{B}(\omega_c,z)|^2}
                   {|\hat{B}_0(\omega_c)|^2\big|_{\Phi=\pi}}\,,
\end{equation}
so that $G = 1$ corresponds to the unperturbed linear-regime response.

\subsection*{Numerical nonlinear Schrödinger model}\label{Methods: simulation}
Observations from the coupled-mode model and the experiment have been compared with those of a nonlinear Schrödinger model. We simulated pulse propagation in a normal-dispersion highly nonlinear fiber (HNLF, length 5\,m) similar to the experimental fiber used for supercontinuum generation around 1550\,nm. The generalized nonlinear Schrödinger equation was solved using a split-step Fourier method in Python (PyTorch version 2.6.0). 

The simulation window spanned $T=50$\,ps with $N=2^{13}$ points (temporal resolution $\Delta t \sim 6$\,fs). Input fields were modelled as Lorentzian pulses to match autocorrelation measurements of the experimental setup, with a full width at half maximum (FWHM) of 180\,fs after compression and a central wavelength of 1550 nm.

The HNLF parameters were defined as follows: The dispersion coefficients were $\beta_{2}=9.3699\times 10^{-28}$
$s^{2}m^{-1}$, $\beta_{3} =  7.6891\times 10^{-42}$ $s^{3}m^{-1}$, $\beta_{4} = 3.1340\times 10^{-55}$ $s^{4}m^{-1}$. The fiber loss was $\alpha = 0.1$  $m^{-1}$. The nonlinear coefficient $\gamma = 10.8\times 10^{-3}$ $W^{-1}m^{-1}$ and the effective mode area $A_{eff}^{-1} = 8\times 10^{10}$ $m^{-2}$. The core radius was $R =  1.92\times 10^{-6}$ $m$. Output spectra and temporal profiles were extracted directly from the simulated fields. 

\subsection*{Correction of Spectral Aberrations}\label{Methods: correction}
To correct for linear and nonlinear aberrations, we use a normalization step that shifts a single phase bit of constant magnitude to estimate the envelope of the spectral response (see Appendix C3 for implementation). A weighting function can then be derived per pixel and applied in post-processing (or pre-processing as an additional frequency-dependent encoding factor). To avoid disproportionate amplification of the baseline, we apply flooring that enforces a minimum weighting value in the peripheral spectral region (Fig. \ref{Figure 3+4}d). The pixel-wise relation between the input and recovered values shows a slight improvement after this correction function is applied, as quantified by the Pearson correlation between inputs and readouts across the spectrum (Fig. \ref{Figure 3+4}e). 

Alternatively, the input pulse shape itself can serve as a weight and be factored out from the output spectrum. This approach is also demonstrated in the Supplementary Information. The nonlinear background could also be mitigated via reference measurements without data imprinted on the pulse; however, in our power and pulse-width regime, this step was not necessary since background contributions were weak. 

\subsection*{Peak signal to noise ratio}\label{Methods: PSNR}
To quantify the fidelity of the reconstructed images relative to the ground-truth MNIST digits, we report the Peak Signal-to-Noise Ratio (PSNR). For two images $x$ (ground truth) and $\hat{x}$(recovered), PSNR is defined as:
\begin{equation}
    PSNR(x,\hat{x})= 10\,log_{10}\left(\frac{1}{MSE(x,\hat{x})}\right)
\end{equation}
where MSE is the mean squared error given by:
\begin{equation}
    MSE(x,\hat{x})= \frac{1}{N} \sum_{i=1}^{N} (x_i - \hat{x_i})^2 \,.
\end{equation}

\section{Declarations}

\noindent

\textbf{Author contributions.} G.R.C. conducted the experiments and simulations, performed data acquisition, analysis, and interpretation, and contributed to manuscript writing and review. M.M. supported the experiments and theoretical modeling and contributed to manuscript review. S.S. supported the experiments and contributed to manuscript review. B.F. provided support in experimental design and methods and contributed to manuscript review. M.C. led the project, designed the experiments, developed the theory and models, supervised the work, supported data analysis and interpretation, and contributed to manuscript writing and review.

\textbf{Funding.}
We acknowledge funding from the Carl Zeiss Foundation through the NEXUS program (project P2021-05-025). 

\textbf{Conflict of Interest.}
The authors declare no conflict of interest.

\textbf{Data availability.}
Data are available via \url{https://doi.org/10.6084/m9.figshare.31441330}.

\textbf{Acknowledgements.}
Special thanks to Nicolas Perron (INRS Montreal/Varennes) for his assistance with simulation implementation. We also acknowledge Julian Hniopek and his team for providing the computing infrastructure needed for this project. We further acknowledge Keysight Germany for access to the real-time scope.

\bibliography{sn-bibliography}

\newpage

\newpage

\newpage

\newpage

\newpage

\newpage

\end{document}